\documentclass[twocolumn]{IEEEtran}

\usepackage{cite}      

\usepackage{graphicx}  
\usepackage{graphicx}

\usepackage{amsmath}   
\begin{document}
%
\title{Design of an Optimal Bayesian Incentive Compatible Broadcast
Protocol for Ad hoc Networks with Rational Nodes}
%
%
\author{N. Rama Suri~~~~~~~~~~~~ Y. Narahari
\thanks{Y. Narahari is a Professor in
the Department of Computer Science and Automation, Indian
Institute of Science, Bangalore, India. E-Mail ID:
hari@csa.iisc.ernet.in. N. Rama Suri is a doctoral student in the
Department of Computer Science and Automation, Indian Institute of
Science, Bangalore. E-Mail ID: nrsuri@csa.iisc.ernet.in. }
\thanks{Manuscript is received on 1st August 2007. Manuscript is revised and submitted on 10th
March 2008.}
\thanks{The
research is supported by ONR Grant No. N00014-06-1-0994. The
research in this paper is partially supported by a research on
Algorithmic Mechanism Design for Complex Game Theoretic Problems
funded by the Office of Naval Research (Grant No.
N0014-06-1-0994), Arlington, VA, USA. We wish to thank Dr Shantanu
Das, Program Manager, Communications and Networking, for the
encouragement and support.}
}

\maketitle


\begin{abstract}
Nodes in an ad hoc wireless network incur certain costs for
forwarding packets since packet forwarding consumes the resources
of the nodes. If the nodes are rational, free packet forwarding by
the nodes cannot be taken for granted and incentive based
protocols are required to stimulate cooperation among the nodes.
Existing incentive based approaches are based on the VCG
(Vickrey-Clarke-Groves) mechanism which leads to high levels of
incentive budgets and restricted applicability to only certain
topologies of networks. Moreover, the existing approaches have
only focused on unicast and multicast. Motivated by this, we
propose an incentive based broadcast protocol that satisfies
Bayesian incentive compatibility and minimizes the incentive
budgets required by the individual nodes. The proposed protocol,
which we call {\em BIC-B\/} (Bayesian incentive compatible
broadcast) protocol, also satisfies budget balance. We also derive
a necessary and sufficient condition for the ex-post individual
rationality of the BIC-B protocol. The {\em BIC-B\/} protocol
exhibits superior performance in comparison to a dominant strategy
incentive compatible broadcast protocol.
\end{abstract}



%
\IEEEpeerreviewmaketitle


\section{Introduction}
Wireless communications industry is currently one of the fastest
growing industries in the world. The industry has several segments
such as cellular telephony, satellite-based communication
networks, wireless LANs, and ad hoc wireless networks.
An {\em ad hoc wireless
network} is an autonomous system of nodes connected through
wireless links. It does {not have any fixed infrastructure}
such as base stations in cellular networks. The nodes in the
network coordinate among themselves for communication. Hence, each
node in the network, apart from being a source or destination, is
also expected to route packets for other nodes in the network.
Such networks find varied applications in real-life environments
such as communication in battle fields, communication among rescue
personnel in disaster affected areas, and wireless sensor
networks.

The conventional protocols for routing, multicasting, and
broadcasting in ad hoc wireless networks assume that nodes follow
the prescribed protocol without any deviation and they cooperate
with one another in performing network functions such as packet
forwarding, etc. However, in many current applications of ad hoc
wireless networks, nodes are {\em autonomous}, {\em rational}, and
{\em intelligent}, and could exhibit strategic behavior. A
wireless node is autonomous because no other wireless node may
have direct control over the decisions or actions taken by
that node. A wireless node is rational in the sense of making
decisions consistently in pursuit of its own objectives. Each
node's objective is to maximize the expected value of individual
payoff measured in some utility scale.
Note that {\em selfishness} or
{\em self-interest} is an important implication of rationality.
A wireless node is
intelligent in the sense that it takes into account its knowledge
or expectation of behavior of other nodes in determining its best response
actions.  The
behavior exhibited by a network of rational and intelligent
nodes can be modeled in a
natural way using {\em game theory\/} \cite{MYERSON97}.
We use the phrase {\em selfish nodes} in this
paper, to refer to rational and intelligent nodes.

There are several recent efforts exploring the use of game
theoretic models in the modeling and analysis of various design
problems in ad hoc networks at different levels of the protocol
stack \cite{VIVEK05}, \cite{GDV04}. It has been applied at (a) the
physical layer level to analyze distributed power control
\cite{CHEN05}, \cite{SARAYDAR02} and waveform adaption; (b) at the
data link layer level to analyze medium access control
\cite{VAIDYA05} and the reservation of bandwidth \cite{FANG04},
\cite{LU04}; and (c) at the network layer level to model the
behavior of the packet forwarding strategies \cite{MOBICOM03},
\cite{MASS04}. Applications at the transport layer and above also
exist, although less pervasive in the literature.

A question of interest in all the situations mentioned above is that of
how to provide appropriate incentives to prevent selfish
behavior by the nodes.
Examples of selfish nodes include: a
node increasing its power without regard for interference it may
cause on its neighbors;  a node immediately retransmitting a frame
in the case of a collision without waiting for the back-off
phrase; a node refusing to forward the transit packets of the
other nodes in the network, etc.
Selfish behavior is generally detrimental to the
overall network performance.
In this paper,
our focus is on studying the
packet forwarding strategies of selfish nodes in the specific
context of broadcast.

\subsection{The Incentive Compatible Broadcast (ICB) Problem}
Broadcast is useful in ad hoc wireless networks in many
situations, for example, route discovery, paging a particular
host, or sending an alarm signal, etc. Successful broadcast
requires appropriate forwarding of the packet(s) by individual
wireless nodes.  The nodes incur certain costs for forwarding
packets since packet forwarding consumes the resources of the
nodes. If the nodes are rational, we cannot take
packet forwarding by nodes for granted.
Reimbursing the forwarding costs or transit costs incurred by the
nodes is a way to make them forward the packets. For this, we need
to know the exact transit costs at the nodes. However, the nodes
may not be willing to reveal the true transit  costs. Broadcast
protocols that induce revelation of true transit costs by the
individual wireless nodes can be called {\em incentive
compatible\/}, following mechanism design terminology. We can
design an incentive compatible broadcast protocol by prescribing an
appropriate allocation rule and payment rule into the broadcast
protocol. We shall refer to the problem of designing such robust
broadcast protocols as the {\em incentive compatible broadcast
(ICB)} problem \cite{SURI06a}. In this paper, we address the ICB
problem in ad hoc wireless networks and offer an elegant solution
using mechanism design theory.


\subsection{Relevant Work}
\label{rel_work}

In the recent times, routing in the presence of selfish nodes has
received significant attention, driven by the need to design
protocols, like routing protocols, multi-cast protocols, etc., for
networks with selfish nodes.

The early research in cooperation of stimulation in ad hoc
wireless networks used a reputation mechanism \cite{CACM00}. These
approaches use techniques such as auditing, use of appropriate
hardware, system-wide optimal point analysis to identify selfish
nodes and isolate non-cooperative nodes from the network. The
watchdog-mechanism in Marti, Giuli, Lai, Baker \cite{MOBICOM00},
the {\em core} mechanism in Michiardi, and Molva \cite{MOLVA02},
the {\em confidant} mechanism in Buchegger, and Boudec
\cite{BUCHEGGER02}, etc. are a few examples of reputation
mechanisms. These methods look for a system-wide optimum point
which may not be individually optimal. Also, the use of hardware
is not always feasible in network settings.

Srinivasan, Nuggehalli, and Chiasserini \cite{INFOCOM03} modeled
the routing situation using the repeated prisoner's dilemma
problem. According to evolutionary game theory, an effective
strategy in this kind of setting is the so-called {\small TIT-FOR-TAT\/}
strategy. In Altman, Kherani, Michiardi and Molva \cite{ALTMAN04},
each node is assumed to forward packets with some probability
which is independent of the source. These models do not take the
dynamics of the network into consideration. A game theoretic model
was introduced by Urpi, Bonuccelli and Giordano \cite{URPI03}
based on static Bayesian games \cite{MYERSON97} to model
forwarding behavior of selfish nodes in ad hoc wireless networks.
Although this model properly formulates the game that the nodes
are playing, it does not allow non-simultaneous decision making.
In addition, the strategies in this framework are not dependent on
past behavior. To get around these problems, Nurmi \cite{NURMI04}
modeled routing in ad hoc wireless networks with selfish nodes as
a dynamic Bayesian game. This model is rich in the sense that it
allows non-simultaneous decision making and incorporating history
information into the decision making process.

Another important approach to designing incentive mechanisms is
based on the techniques of microeconomics. Mechanism design is a
powerful tool to model such situations. There have been several
efforts to design incentive mechanisms for routing in ad hoc
wireless networks in the presence of selfish nodes. Feigenbaum,
Papadimitriou, Sami, and Shenker \cite{PODC02} and Hershberger and
Suri \cite{FOCS01} developed an incentive mechanism to address the
truthful low cost routing (unicast) problem. The model consists of
$n$ nodes where each node represents an autonomous system. They
assume that each node $k$ incurs a transit cost $c_k$ for
forwarding one transit packet. For any pair of nodes $i$ and $j$
of the network, $T_{i,j}$ is the total amount of traffic
originating from $i$ and destined for node $j$. The payments to
nodes are computed using the Vickrey-Clarke-Groves (VCG) payment
rules \cite{MASCOLELL95,GARG-MD06}.  The authors in \cite{PODC02}
and \cite{FOCS01} presented a distributed method such that each
node $i$ can compute its payment $p_{i,j}^{l}$ to node $l$ for
carrying the transit traffic from node $i$ to node $j$ if node $l$
is on the least cost path from $i$ to $j$.

A similar type of model was presented by Anderegg and Eidenbenz
\cite{MOBICOM03} for stimulating cooperation among selfish nodes
in ad hoc networks using an incentive scheme. This model
generalizes one aspect of \cite{PODC02} by associating several
costs to each node, one per each neighbor, instead of just one.
This leads to a model based on {\em edge weighted graph}
representation of the ad hoc network. The VCG mechanism is used to
compute a power efficient path, where each node determines the
power level required to transit/forward the packets. A node by
itself cannot determine its power level required because it needs
feedback information in the form of packets from its neighbors. As
the nodes are selfish and non-cooperative, this feedback
information may allow a node to cheat its neighbors in order to
raise its own payoff. The authors of \cite{MOBICOM03} did not
address this issue. Eidenbenz, Santi, and Resta \cite{IPDPS05}
modified the model in Anderegg and Eidenbenz \cite{MOBICOM03} by
using the VCG mechanism to compute the payments to the nodes, but
the sender is charged the total declared cost of a second least
cost path, that is the least cost path with all nodes in the cost
efficient path removed. This requires the existence of at least
two node disjoint paths between the sender and the receiver. Wang
and Li \cite{WMAN04} proposed strategy-proof mechanisms for the
truthful unicast problem. They  also presented an algorithm for
fast computation of payments to nodes and a distributed algorithm
for payment computation.

Zhong, Li, Liu, and Yang \cite{LILI05} used a two-stage routing
protocol to model the routing problem in ad hoc wireless networks
with selfish nodes. They integrated a novel cryptographic
technique into the VCG mechanism to solve the link cost dependence
problem. Zhong, Chen, and Yang \cite{ZHONG-INFOCOM03} proposed a
system called Sprite, which combines incentive methods and
cryptography techniques to implement a group cheat-proof ad hoc
routing system. Lu, Li, Wu \cite{JIEWU06} embed an
incentive-compatible, efficient, and individual rational payment
scheme into the routing protocol in ad hoc networks which consist
of selfish nodes. Unlike traditional routing protocols in ad hoc
networks, which only elicit cost information from selfish nodes,
this model motivates selfish nodes to report truthfully both their
stability and cost information.

\begin{figure}[h]
\begin{center}
\rotatebox{360}{\scalebox{1.0}{\includegraphics{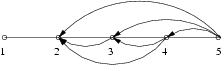}}}
\caption{\label{uni_broad} Relating incentive compatible unicast
solution to the {\em ICB} problem}
\end{center}
\end{figure}

In all the above solution approaches for the incentive compatible
unicast problem (also known as the {\em truthful unicast problem}),
the intermediate nodes on the path between the source node and the
destination node are compensated for forwarding the packet. Let us
apply this technique of service reimbursement to the incentive
compatible broadcast ({\em ICB}) problem and see the consequences.
Consider a portion of the network as shown in Figure
\ref{uni_broad}. Let us consider node $1$ to be the source of
broadcast. Nodes $2$, $3$, $4$, and $5$ are the intended
destinations of the broadcast packet. Node $5$ needs to reimburse
the nodes $2$, $3$, and $4$, since these nodes are intermediate
nodes on the path between node $1$ and node $5$ and they forward
the broadcast packet. For similar reasons, node $4$ needs to
reimburse the nodes $2$ and $3$ and node $3$ needs to reimburse
node $2$. On the whole, node $2$ receives payments from nodes $3$,
$4$, and $5$ for forwarding a packet once. Similarly, node $3$
receives payments from node $4$, and node $5$. Hence we end up
with high values of over payments leading to an inefficient
solution to the {\em ICB} problem. It is important to observe, in
the context of broadcast, that all the nodes in the network are
intended destinations and there is no concept of intermediate
nodes as in the case of unicast.

\input{epsf}

Wang and Li \cite{MOBICOM04}, \cite{MASS04} have proposed
strategy-proof mechanisms for the truthful multicast problem in ad
hoc wireless networks with selfish nodes. The authors, in this
model, assume that the nodes in the multicast set forward the
packets for free among themselves. However, this assumption may
not be credible in real world applications, especially when the
nodes are rational. In our work, we assume that nodes, being
rational, do not
necessarily forward the broadcast packets for free.
This assumption clearly captures the real-world more
accurately.

Suri and Narahari \cite{SURITECH07,SURI07} proposed a dominant strategy
incentive compatible mechanism, DSIC-B,
that is built into the corresponding broadcast
protocol, as a solution to the ICB problem.
In the {\em DSIC-B} protocol, a node pays only for the node from which it has
received the broadcast packet and these payments are designed such
that the mechanism is truthful in the dominant strategy sense, that is,
true cost revelation is a best response for each node irrespective of
what the other nodes report.
There are, however, two limitations of using the {\em DSIC-B\/}
protocol \cite{SURITECH07,SURI07}.
\begin{itemize}
\item {\bf (L1)}: The underlying graph of ad hoc wireless network
must be bi-connected. \item {\bf (L2)}: The {\em DSIC-B\/}
protocol is {\em not budget balanced} \cite{MASCOLELL95,GARG06}.
\end{itemize}
The limitation {\bf (L1)} implies that the DSIC-B
protocol works only for networks where
the underlying graph of that network is bi-connected. This means
that there cannot exist any cut vertex in the underlying graph. As
the number of nodes in the network increases, sustaining this
condition is difficult.
The limitation {\bf (L2)} implies that certain external budget
needs to be injected into the network to sustain the
running of the protocol.


To summarize the relevant work,
\begin{itemize}
\item Several game theoretic models and mechanism design solutions have been
proposed in the context of routing and other functions in ad hoc networks.
However, most of these address only unicast and multicast.
\item The mechanism design solutions proposed in the literature are all
based on the VCG mechanisms. Though the VCG mechanisms guarantee the
extremely desirable property of dominant strategy incentive compatibility,
the incentive budgets required tend to be very high.
In fact, there is a recent study
\cite{PATRICK06} on the limitations of VCG mechanisms (i.e.
dominant strategy incentive compatible mechanisms) in the design
of protocols for networks with selfish nodes.
\item There are not too may studies in the literature which address
the incentive compatible broadcast problem. Also the existing approaches
for incentive compatible broadcast have many limitations. Further,
the existing solutions available for unicast and multicast invariably
lead to inefficient solutions when applied to the broadcast
setting.

\end{itemize}

Till this point, we have set the stage for the need of designing
efficient incentive based mechanisms for ICB problem. We propose
to use Bayesian model to design such incentive mechanisms. In the
following section, we argue in the favor of choosing a Bayesian
model with which we can overcome several limitations due to VCG
based mechanisms.

\subsection{Need for a Bayesian Model}
There are several reasons why the Bayesian model
makes more sense than a VCG type of model:
\begin{itemize}
 \item The payments in the VCG model are almost always much higher
       when compared with the Bayesian models (This is natural because
       the dominant strategy incentive compatibility property guaranteed
       by the VCG model is a much stronger property and hence entails
       higher incentives to be paid). This is well known in the mechanism design literature.
 \item In the case of the Bayesian model, budget balance is satisfied.
       On the other hand, in the case of the VCG model, budget balance is difficult to
        achieve, except under very special settings. Budget balance
       is a desirable property because it
       ensures that the protocol does not require any external source of funding
       and therefore is self-sustaining.
  \item In the case of broadcast, the context is such that
there is one source node and the rest of the nodes are receivers
of the broadcast packet. Since the nodes are rational, they may
need to make payments to receive the broadcast packet(s). It makes
sense to expect all the nodes that receive packets to make the same
payments. This is because the source node has no way of distinguishing
among the remaining nodes (which are all receivers). We have shown in
this paper that the payments as computed by the Bayesian model
will be identical for all the receiving nodes. Thus the Bayesian
model captures the real-world in a natural way.
\end{itemize}
However, there are two issues with the use of Bayesian approach.
First, due to the high payments entailed by VCG-based solutions,
there is a compelling need to look for Bayesian incentive
compatibility, which is a much weaker form of incentive
compatibility. Second, we may lose out on individual rationality.
We have however derived a sufficient condition under which
individual rationality is also guaranteed by our Bayesian model.


\subsection{Contributions and Outline of the Paper}
Motivated by the above considerations, we offer the following
contributions in this paper.
\begin{itemize}
\item We propose an incentive based
broadcast mechanism that satisfies Bayesian incentive compatibility
and minimizes the incentive budgets required by the individual nodes.
Bayesian incentive compatibility ensures that truth revelation is
a best response for each node whenever all other nodes are also
truthful.

\item The above proposed mechanism, which we call {\em BIC-B\/} (Bayesian incentive
compatible broadcast) mechanism,
also satisfies budget balance. This ensures that the protocol is self-sustaining
and does not require any external budget for sustaining the running
of the protocol.

\item We also derive a necessary and sufficient condition
for the ex-post individual rationality of the BIC-B mechanism.
Individual rationality guarantees that the nodes do not incur
negative payoffs by participating in the protocol.

\item We present an approach for a protocol implementation
based on the BIC-B mechanism. We also explain how the payments to the nodes are
computed following the proposed mechanism.

\item We show that the {\em BIC-B\/} protocol exhibits superior performance
in comparison to existing protocols for incentive compatible broadcast
such as DSIC-B.
\end{itemize}


The rest of the paper is organized as follows.
In Section II, we present a game theoretic model for the incentive compatible
broadcast problem and offer a Bayesian incentive compatible mechanism design
solution for the problem. In Section III, we investigate different properties of the
proposed BIC-B protocol such as budget balance, budget minimization, and
individual rationality. In Section IV, we present a simulation experiment to
exhibit the superior performance of the BIC-B protocol. Section V concludes the paper and
outlines a few directions for future work.


\section{The Model and the BIC-B Protocol}
\label{the_model} An ad hoc wireless network, in this paper, is
represented by an undirected graph $G = (N,E)$, where $N = \{1, 2,
...,n\}$ is the set of $n$ wireless nodes and $E$ is the set of
communication links between the nodes. There exists a
communication link between two nodes $i$ and $j$, if a node $i$ is
reachable from node $j$, and node $j$ is also reachable from node
$i$. Thus we have an undirected graph representation. We assume
that wireless nodes use directional antennas and a single
transmission by a node may be received by only a subset of nodes in its
vicinity. We note that all nodes in the graph $G$ are connected.

We assume that the nodes in the ad hoc wireless network are owned
by rational and intelligent individuals and so their objective is
to maximize their individual goals. For this reason, they may not
always participate loyally in key network functions, such as
forwarding the packets, since such activity might consume the
node's resources, such as battery power, bandwidth, CPU cycles, etc.
Let each node $i$ incur a cost $\theta_i$ for forwarding a packet.
For simplicity, we assume that $\theta_i$ is independent of the
neighbor from which the packet is received and the neighbor to
which the packet is destined. We can represent $\theta_i$ as the
weight of node $i$ in the graph $G$. This implies that we work
with a {\em node weighted graph}.

Consider the task of broadcast in such a setting. Assume that the
source of the broadcast is node $s$. Not all remaining nodes
may be connected to node $s$, hence appropriate intermediate
nodes have to forward
the broadcast packet to ensure that the packet reaches all the
nodes in the network. As explained above, the nodes that forward
the packet(s) incur costs. This means, we need to look at a tree
that spans all the nodes and has the source node $s$ as the root
of the tree. We call such a tree as {\em Source Rooted Broadcast
Tree (SRBT)}. Given an {\em SRBT}, we design an
appropriate incentive scheme or mechanism that is built into the broadcast
protocol.

%
\subsection{A Game Theoretic Model}
A game theoretic model in the above setting is described
below.
\begin{itemize}
\item There are $n$ wireless nodes, $1,2,\ldots,n$ in the
ad hoc network. Let
$N=\{1,2,\ldots,n\}$.
\item Each wireless node $i$ privately observes a signal
$\theta_i$ that determines the cost for the node $i$ to forward a
packet. The value of $\theta_i$ is known to agent $i$
deterministically since $\theta_i$ is dependent of the consumed
CPU cycles, battery power, bandwidth, etc. However this value is
not known to other nodes. Hence we call $\theta_i$ as the {\em
private value} or {\em type} of node $i$.
\item We denote by $\Theta_i$ the set of types of node $i$,
$i=1,2,\ldots,n$. $\Theta=\times_{i \in N}\Theta_{i}$ is the set
of all type profiles of the nodes. $\theta=(\theta_1, \theta_2,
\ldots, \theta_n)$ represents a typical type profile of the
nodes.
\item It is assumed that the types of the nodes are drawn from a
common prior distribution $\varphi$. We make the standard
assumption that
individual belief functions $p_i$ are computed using the above
common prior. The belief
function $p_i$ ($i=1,2, \ldots,n$) describes
the belief of node $i$ about the types of the remaining nodes.
\item $X$ is the set of different possible outcomes. Each outcome
specifies {\em the set of routers} and {\em payments} to those
routers for forwarding packets. The selection of a particular
choice of set of routers depends on the type profile
$\theta=(\theta_1, \theta_2, \ldots, \theta_n)$ of the nodes. This
is captured by social choice function $f(.)$
\cite{MASCOLELL95}.
\item If a node $i$ forwards a packet, it incurs a cost $\theta_i$
and hence it must be compensated appropriately. More generally, we
capture this notion by a utility function $u_{i}:\; X \times
\Theta_i \rightarrow R$. In particular, we assume that the
utility functions are {\em quasi-linear\/}.
\end{itemize}
Following the terminology of mechanism design \cite{MASCOLELL95},
a mechanism $M=(S_{1},S_{2},\ldots,S_{n},g(.))$ is a collection of
strategy sets $S_{1},S_{2}, \ldots, S_{n}$ and an outcome function
$g(.)$. Here $S_{i}$ captures the possible announcements of node
$i$ regarding its incurred costs. The outcome function $g(.)$ is
defined as $g: \; S_{1} \times S_{2} \times \ldots \times S_{n}
\rightarrow X$. A mechanism is a framework which prescribes an
action set for each player and specifies how these action profiles
are transformed into outcomes. The outcome function $g(.)$ gives
the rule for obtaining outcomes from action profiles. A mechanism
induces a Bayesian game $(N,(S_{i}),(\Theta_{i}),(p_{i}),(U_{i}))$
where $U_{i}(.)$ is computed from $u_{i}(g(.),\theta_{i})$ \cite{MASCOLELL95}.
This induced Bayesian game can have a solution which is either a
dominant strategy equilibrium or a Bayesian Nash equilibrium.
Accordingly we have either a dominant strategy incentive
compatible mechanism or a Bayesian incentive compatible mechanism,
respectively.

\subsection{The BIC-B Protocol}
\label{bic-b4}
%


We now present a mechanism that implements the social
choice function (SCF) $f(\theta) = (k(\theta),t_1(\theta), \ldots,
t_n(\theta)),$ $\forall \theta \in \Theta$. Here $k(\theta)$, and
$t_i(\theta), \; \forall i \in N$ are interpreted in the following
way. $k(\theta)$ is the allocation rule that represents which
nodes in the network have to forward the packet, given the profile
$\theta$ of types. The vector $(t_1(\theta), \ldots, t_n(\theta))$ gives
payments received by the nodes, given the profile
$\theta$ of types. For any $i \in N$, if $t_i(\theta) > 0$, then
the interpretation is that $i$ receives some positive amount and
if $t_i(\theta) < 0$, then the interpretation is that $i$ pays
some positive amount.

Assume that we are given the $SRBT$ corresponding to the graph
under consideration. We design, based on the given {\em SRBT}, an
the following payment scheme that determines the payments
$(t_i(\theta))_{i \in N}$ to the individual nodes for a broadcast.
In the {\em SRBT}, all the internal nodes
forward the broadcast packet. We call such packet forwarding nodes
as {\em routers}, and represent the set of routers by $R$. Note that
each outcome of the SCF $f(.)$ has an allocation rule and a payment
rule. We define the allocation rule in the following way:
$\forall \theta \in \Theta$, $\forall i \in N$,

\begin{center}
\begin{tabular}{l c l}
$k_{i}(\theta)$ & = & $1, \quad$ if $ i \in R$ \\
                & = & $0, \quad$ if $ i \notin R$
\end{tabular}
\end{center}

The valuation function, $v_i(k(\theta),\theta_i)$, of node $i$ is
its cost to forward a transit packet. From the allocation rule
$k(.)$ in our SCF $f(.)$, we get, $\forall \theta \in \Theta$,
$\forall i \in N$,

\begin{eqnarray}
v_i(k(\theta),\theta_i) & = & - \theta_i,  \quad {\rm if} \; i \in R \label{val_r} \\
                        & = & 0, \quad  {\rm if} \;  i \notin R
\label{val_nr}
\end{eqnarray}

The broadcast packets from the source node travel through the
paths specified in the {\em SRBT}. To compensate the incurred cost
of the routers in the network, we need to determine payments to
the nodes. We follow the payment rule of the
classical {\em dAGVA mechanism} or {\em
expected externality mechanism} \cite{MASCOLELL95,DGARG08}, to
compute the payments to the nodes in our scheme. Using the payment
rule of the dAGVA mechanism \cite{MASCOLELL95,DGARG08}, $\forall i \in N$,
$\forall \theta \in \Theta$, we get

\begin{center}
\begin{tabular}{l c l}
$t_i(\theta)$ & = & $E_{{{\theta}}_{-i}}\left[\sum_{l \not =
i}v_l(k(\theta),{{\theta}}_l)\right]$\\
              &   & $ - \left(\frac{1}{n-1}\right) \sum_{j \neq i}
E_{{{\theta}}_{-j}}\left[\sum_{l \not =
j}v_l(k(\theta),{{\theta}}_l)\right]$
\end{tabular}
\end{center}

From (\ref{val_r}), (\ref{val_nr}) we get, $\forall i \in N$,
$\forall \theta \in \Theta$,

\begin{equation}
t_i(\theta)=\left(\frac{1}{n-1}\right) \sum_{j \neq i}
E_{{{\theta}}_{-j}}\left[\sum_{l \in R, \: l \not = j}
\theta_{l}\right] -  E_{{{\theta}}_{-i}}\left[\sum_{l \in R, \: l
\not = i}\theta_{l}\right] \label{payment_rule}
\end{equation}

where $E_{{{\theta}}_{-i}}\left[\sum_{l \in R, \: l \not =
i}\theta_{l}\right]$ is interpreted as the total expected value to
node $i$ that would be generated by all the remaining nodes in the
absence of node $i$. This completes the characterization of the
payment rule of the {\em BIC-B\/} mechanism. Note that this
mechanism with the above payment rule is incentive compatible.
This directly follows from the incentive compatibility property of
the dAGVA mechanism \cite{MASCOLELL95,DGARG08}.

In the following, we first present an illustrative example to
understand the details of the proposed payment scheme and we then
investigate the properties of the mechanism.

\subsection{An Example}
\begin{figure}[h]
\begin{center}
\rotatebox{360}{\scalebox{1.0}{\includegraphics{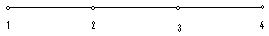}}}
\caption{\label{ie2_graph} Illustrative example 1}
\end{center}
\end{figure}

Now we provide an example to illustrate
the payment scheme in {\em BIC-B} protocol.
Let us consider the graph in Figure
\ref{ie2_graph}, which is {\em not bi-connected}. We recall that
the types of nodes are their incurred transit costs. We consider
the type sets of nodes as $\Theta_1=\{10,11\}$,
$\Theta_2=\{15,16\}$, $\Theta_3=\{12,13\}$, $\Theta_4=\{7,8\}$.
Now let us assume that the players belief probability functions
are independent discrete uniform distributions with equal
probabilities for all types. and $\theta = (10, 15, 13, 8)$ is the
announced cost profile.

\input{epsf}


In this example, $SRBT$ is also the same the original graph. Now,
the allocation rule is $k(\theta) = (0, 1, 1, 0)$. We note that
$N=\{1, 2, 3, 4\}$ and $R=\{2, 3\}$. The valuation functions of
nodes are, $v_1(k(\theta))=0$, $v_2(k(\theta))=-15$,
$v_3(k(\theta))=-13$, $v_4(k(\theta))=0$. Now we compute the
payments using the payment rule (\ref{payment_rule}) of {\em
BIC-B} protocol.

The payment computation for node $1$:\\
\begin{tabular}{l c l}
$t_1(\theta)$ & = & $ \left(\frac{1}{4-1}\right) \sum_{j \neq 1}
E_{{{\theta}}_{-j}}\left[\sum_{l \in R, \: l \not = j}
\theta_{l}\right]$ \\
              &   & $-  E_{{{\theta}}_{-1}}\left[\sum_{l \in R, \:
l \not = 1}\theta_{l}\right] $ \\
              & = & $\left(\frac{1}{3}\right) E_{{{\theta}}_{-2}}\left[\sum_{l \in R, \: l \not = 2}
\theta_{l}\right]$ \\
              &   & $+ \left(\frac{1}{3}\right)
E_{{{\theta}}_{-3}}\left[\sum_{l \in R, \: l \not = 3}
\theta_{l}\right]$ \\
              &   & $ + \left(\frac{1}{3}\right)
E_{{{\theta}}_{-4}}\left[\sum_{l \in R, \: l \not = 4}
\theta_{l}\right]$ \\
              &   & $ - E_{{{\theta}}_{-1}}\left[\sum_{l \in R,
\: l
\not = 1}\theta_{l}\right]$ \\
              & = & $\left(\frac{1}{3}\right) \left[
E_{{{\theta}}_{-2}}\left[\theta_{3}\right] +
E_{{{\theta}}_{-3}}\left[\theta_{2}\right] +
E_{{{\theta}}_{-4}}\left[\theta_{2} + \theta_{3} \right] \right]$
\\
              &   & $ - E_{{{\theta}}_{-1}}\left[\theta_{2} + \theta_{3}\right]$ \\
              & = & $ \left(\frac{1}{3}\right) \left[
E_{{{\theta}}_{-2}}\left[\theta_{3}\right] +
E_{{{\theta}}_{-3}}\left[\theta_{2}\right] \right]$ \\
              &   & $+ \left(\frac{1}{3}\right) \left[ E_{{{\theta}}_{-4}}\left[\theta_{2}\right] +
E_{{{\theta}}_{-4}}\left[\theta_{3} \right]\right]$ \\
              &   & $ - \left[ E_{{{\theta}}_{-1}}\left[\theta_{2}\right] + E_{{{\theta}}_{-1}}\left[\theta_{3}\right] \right]$\\
              &  & (since types are statistically independent)\\
              & = & $ \left(\frac{1}{3}\right) \left[
12.5 + 15.5 + 15.5 + 12.5 \right]$\\
              &   & $ - \left[15.5 + 12.5 \right]$\\
              & = & $-9.33$
\end{tabular}


In the similar fashion, we can compute the payments to the
remaining nodes also. The payments to nodes are:
$t_1(\theta)=-9.3$, $t_2(\theta)=11.3$, $t_3(\theta)=$7.3,
$t_4(\theta)=-9.3$. In this example, node $1$ and node $4$ do not
forward the packets. Observe that they pay the same amount, namely
$9.33$. Since node $2$ and node $3$ are routers, they receive
amounts $11.33$ and $7.33$ respectively. Further observe that
$\sum_{i=1}^{4}t_i(\theta) = 0$.

\section{Properties of the {\em BIC-B} Protocol}
\label{properties}
\subsection{Budget Balance}
\begin{figure}[h]
\begin{center}
\rotatebox{360}{\scalebox{0.4}{\includegraphics{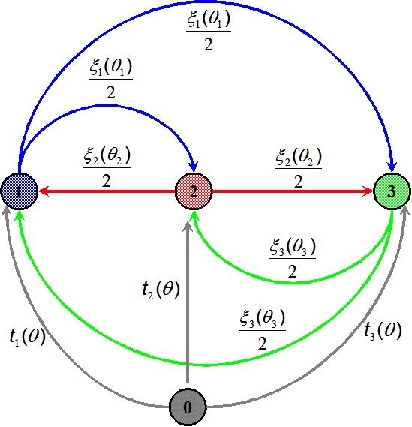}}}
\caption{\label{dagva} An interpretation of the payment rule}
\end{center}
\end{figure}

Assume that $\xi_i(\theta_i) =
E_{{{\theta}}_{-i}}\left[\sum_{l \in R, \: l \not =
i}\theta_{l}\right]$, $\forall i \in N$.
To show budget balance, we need to show that the sum of the
payments received and the payments made by the nodes in the
network is zero, i.e. $\sum_i t_i(\theta) = 0, \; \; \forall
\;\theta \in \Theta$. Let us assume that $\xi_i(\theta_i) =
E_{{{\theta}}_{-i}}\left[\sum_{l \in R, \: l \not =
i}\theta_{l}\right]$, $\forall \theta \in \Theta, \; \forall i \in
N$. Then, from (\ref{payment_rule}), we get $\forall \theta \in
\Theta$
\begin{eqnarray*}
t_i(\theta)&=&\xi_i(\theta_i) - \left(\frac{1}{n-1}\right)\sum_{j\not=i}\xi_j(\theta_j) \\then, \\
\sum_{i=1}^{n}t_i(\theta)&=&\sum_{i=1}^{n}\xi_i(\theta_i) -  \left(\frac{1}{n-1}\right) \sum_{i=1}^{n}\sum_{j\not=i}\xi_j(\theta_j) \\
&=&\sum_{i=1}^{n}\xi_i(\theta_i) -  \left(\frac{1}{n-1}\right)
\sum_{i=1}^{n}(n-1)\xi_j(\theta_j)
\end{eqnarray*}
\begin{center}
$\Rightarrow \qquad \sum_{i=1}^{n}t_i(\theta)=0$
\end{center}
It can be noted that
each node $i$ distributes
$\xi_i$ equally among the remaining $(n-1)$ nodes.
Figure
\ref{dagva} shows this interpretation for a graph with 3 nodes.

\input{epsf}


Let us represent the flow into the node with positive sign and the
flow going out from the node by negative sign. Now the payment,
$t_1(\theta)$, for node $1$ can be written as

\begin{center}
\begin{tabular}{l c l}
$t_1(\theta_1,\theta_2,\theta_3)$ & = & $\frac{1}{2}
\xi_{2}(\theta_2) + \frac{1}{2} \xi_{3}(\theta_3) -
\xi_{1}(\theta_1)$ \\
\end{tabular}
\end{center}
Similarly, we can write down the payments to the remaining two nodes.

\subsection{Payments by Non-Router Nodes}
Recall that $N$ is the set of nodes and $R$
represents the set of routers in the network. We now state  {\em Lemma
1} which is useful in proving {\em Lemma 2} and subsequently {\em Theorem 1}.
\vspace{0.3cm}\\
\textbf{Lemma 1:} For any $i \in R$ and for any $j \notin R$, we
have
$  E_{{{\theta}}_{-i}}\left[\sum_{l \in R, \: l \not =
i}\theta_{l}\right] $  = $ E_{{{\theta}}_{-j}}\left[\sum_{l \in R,
\: l \not = i}\theta_{l}\right]$.\\
\noindent
\textbf{Proof:} Proof is provided in the
appendix. \\[2mm]
\noindent \textbf{Lemma 2:} For the {\em BIC-B} protocol,
$t_i(\theta) < 0$, $\forall i \notin R, \; \forall \theta \in
\Theta$.
That is, {\em the nodes other than the routers will pay a positive
amount of money for receiving the packet(s)}.
\vspace{0.1cm}\\
\noindent
\textbf{Proof:} If $R$ is empty, then the source node can reach
all the remaining nodes in the network within a single hop. In
that case, the payments will be 0. We are not interested in such a
trivial situation. So we assume that,
\begin{equation}\label{**}
|R| > 0
\end{equation}
Let us assume that, $\forall j \notin R$,
\begin{center}
\begin{tabular}{l c l}
$\Gamma_j $ & = &  $ E_{{{\theta}}_{-j}}\left[\sum_{l \in R, \: l \not = j}\theta_{l}\right] $ \\
                       & = &  $ E_{{{\theta}}_{-j}}\left[\sum_{l \in R}\theta_{l}\right]
                       \qquad$ (since $j \notin R$)
\end{tabular}
\end{center}
Since the types of nodes are statistically independent, the values
of $\Gamma_j$, $\forall j \notin R$, are all the same. We
represent this value with $\Gamma$. That is,
\begin{equation}
\Gamma = \Gamma_j, \qquad \forall j \notin R. \label{nr_equality}
\end{equation}
Now, let us assume that   $\Upsilon_i =
E_{{{\theta}}_{-i}}\left[\sum_{l \in R, \: l \not =
i}\theta_{l}\right] $, $\forall i \in R$. Then
\begin{center}
\begin{tabular}{l c l}
$\Upsilon_i $  & = & $  E_{{{\theta}}_{-i}}\left[\sum_{l \in R, \: l \not = i}\theta_{l}\right] $   \\
                       & = &  $ E_{{{\theta}}_{-j}}\left[\sum_{l \in R, \: l \not = i}\theta_{l}\right] \; $ \\
                       &   &  (consequence of Lemma 1) \\
                       & $<$ &  $ E_{{{\theta}}_{-j}}\left[\sum_{l \in R}\theta_{l}\right]$ $\qquad$ (since $i \in R$) \\
                       & = & $\Gamma$ $\qquad$ (from equation (\ref{nr_equality}))
\end{tabular}
\end{center}
So, we can conclude that
\begin{equation}
      \Upsilon_i  <  \Gamma,  \qquad  \forall i \in R.
      \label{major}
\end{equation}
From the payment rule (\ref{payment_rule}) of the {\em BIC-B}
protocol, we have $\forall i \notin R$,
\begin{center}
\begin{tabular}{l c l}
$t_i(\theta)$ & = & $ \left(\frac{1}{n-1}\right) \sum_{j \neq i,
\; j \in N } E_{{{\theta}}_{-j}}\left[\sum_{l \in R, \: l \not =
j} \theta_{l}\right]$ \\
              &   & $- E_{{{\theta}}_{-i}}\left[\sum_{l \in R, \; l \not = i}\theta_{l}\right]$ \\
              & = & $ \left(\frac{1}{n-1}\right) \sum_{j \in R} E_{{{\theta}}_{-j}}\left[\sum_{l \in R, \: l \not = j}
              \theta_{l}\right]$ \\
              &   & $+ \left(\frac{1}{n-1}\right) \sum_{j \neq i, \; j \notin R} E_{{{\theta}}_{-j}}\left[\sum_{l \in R, \: l \not = j} \theta_{l}\right]$ \\
              &   & $ - E_{{{\theta}}_{-i}}\left[\sum_{l \in R, \: l \not = i}\theta_{l}\right]$\\
              & = & $ \left(\frac{1}{n-1}\right) \sum_{j \in R} \Upsilon_j + \left(\left(\frac{1}{n-1}\right) \sum_{j \neq i, \; j \notin R}\Gamma \right) - \Gamma$ \\
              &   & (since from equation (\ref{nr_equality})) \\
              & = &  $ \left(\frac{1}{n-1}\right) \sum_{j \in R} \Upsilon_j$  $ + \left(\frac{|N|-|R|-1}{n-1} - 1 \right) \Gamma$ \\
              & = &  $\left(\frac{1}{n-1}\right) \sum_{j \in R} \Upsilon_j$  $- \left(\frac{1}{n-1}\right) \sum_{j \in R} \Gamma$ \\
              & = &  $\left(\frac{1}{n-1}\right) \sum_{j \in R} (\Upsilon_j - \Gamma)$ \\
              & $<$ &  0, $\qquad$ (from (\ref{major})).
\end{tabular}
\end{center}

According to our previous interpretation, $t_i(\theta) < 0$ means
that node $i$ needs to pay the specified amount. This completes the proof the lemma. ({\em Q.E.D.}). \\[2mm]
\noindent \textbf{Observation 1:} From the proof of Lemma 2, we
know that $\forall i \notin R$,
\begin{center}
$t_i(\theta)$ = $\left(\frac{1}{n-1}\right) \sum_{j \in R}
(\Upsilon_j - \Gamma)$
\end{center}
Note that the right hand side of the above expression is
independent from $i$. Hence, using the {\em BIC-B} protocol,
$t_i(.)$, $\forall i \notin R$ are all the same. The immediate
implication is that the payments made by the nodes other than the
routers are the same.

\subsection{Optimality of the {\em BIC-B} Payments}
Here we prove the optimality of the payments prescribed by the
{\em BIC-B} protocol. We note that an appropriate allocation rule
is employed to determine the SRBT of the underlying graph of the
ad hoc wireless network for the broadcast task. We define {\em
cost of SRBT} is sum of the forwarding costs of the router nodes.
It is the case that any allocation rule tries to minimize the cost
of the SRBT. If it is hard to find optimal SRBT, then we assume
that an appropriate approximation algorithm is used to determine
the SRBT.

We make the following observation for the sake of Theorem 1.\\
{\bf Observation 2:} Consider two type profiles $\theta$ and
$\theta^{'}$. Assume that these two types are different only with
respect to the type of node $i$. Using the BIC-B mechanism, when
the types are $\theta$ and $\theta^{'}$, the payments to node $i$
are $t_i(\theta)$ and $t_i(\theta^{'})$ respectively. Note that if
the corresponding SRBT is the same for both the types $\theta$ and
$\theta^{'}$, then the set of routers is the same and hence the
payments $t_i(\theta)$ and $t_i(\theta^{'})$ respectively to node
$i$ are the same. This is because in the payment rule
(\ref{payment_rule}) of BIC-B mechanism, the quantities involved
only look for expected values of the types of the routers.
\vspace{0.2cm}\\[1mm]
\noindent {\bf Theorem 1:} {\em For the given {\rm SRBT} structure
of the underlying graph $G$ of the ad hoc wireless network, the
payment to any node using the {BIC-B\/} mechanism is minimum among
all other Bayesian incentive compatible mechanisms based on
SRBT\/}.
\vspace{0.2cm}\\
\noindent {\bf Proof:} We provide a contradiction to prove the
statement. Let $t(.)$ be the payment rule of the {\em BIC-B\/}
mechanism based on the given SRBT. Assume that there exists
another Bayesian incentive compatible mechanism with payment rule
$\hat t(\cdot)$ such that the payment to a router node $i$ is
strictly less. That is,
\begin{equation}
\hat t_i(\theta) < t_i(\theta), \qquad \forall \theta \in \Theta,
\;  {\rm for \;\; some\/}\;\; i \in N  \label{relation1}
\end{equation}

We construct a contradiction to show that $\hat t(\cdot)$ is not
incentive compatible for the node $i$ under some cost profile
$\theta^{'}$. We construct $\theta^{'}$ from $\theta$ by replacing
the cost $\theta_i$ of node $i$ with $(t_i(\theta)+\epsilon)$,
where $\epsilon >0$. That is, if the cost profile $\theta$ is such
that $\theta = (\theta_i, \theta_{-i})$, (where the $n$-tuple
$(\theta_i, \theta_{-i})$ indicates a cost profile of the nodes
where the cost of node $i$ is $\theta_i$ and the costs of
remaining nodes is represented by $\theta_{-i}$), then the cost
profile $\theta^{'} $ is such that $\theta^{'} =
((t_i(\theta)+\epsilon), \theta_{-i})$. Recall that node $i$ is a
router in the corresponding SRBT when type profile is $\theta$. We
now consider the following two cases.

{\em Case 1:} Consider $\theta^{'}$ is the announced cost profile.
Assume that node $i$ is a router in the corresponding SRBT. This
assumption is true definitely if $(t_i(\theta)+\epsilon) <
\theta_i$ because node $i$ is a router even with the type
$\theta_i$. That is, the SRBT is one and the same under both the
type profiles $\theta$ and $\theta^{'}$. Using the payment rule
$\hat t(.)$, the payment to node $i$ is $\hat t_i(\theta^{'})$.
Now the gain from being a router to the node $i$ is $\hat
t_i(\theta^{'}) - (t_i(\theta)+\epsilon) < t_i(\theta^{'}) -
(t_i(\theta)+\epsilon) = - \epsilon < 0$ (using (\ref{relation1})
and Observation 2).

{\em Case 2:} Consider $\theta^{'}$ is the announced cost profile.
Assume that node $i$ is not a router in the appropriate SRBT. Then
there is no forwarding cost incurred to node $i$. Then there is no
issue.

Hence for node $i$, if its type is $(t_i(\theta)+\epsilon) <
\theta_i$ and then it happens to be a router in the corresponding
SRBT and it gets negative gains. Due to the appropriate arguments
in Theorem 2, there exists a $\theta_i \in \Theta_i$ such that
$(t_i(\theta)+\epsilon) < \theta_i$ is true. Hence the mechanism
with $\hat t(\cdot)$ as the payment rule is not incentive
compatible. This provides the required contradiction. ({\em
Q.E.D.}).

\subsection{Individual Rationality of the {\em BIC-B} Protocol}
We now investigate the {\em individual rationality (IR)} of the
{\em BIC-B} protocol. In particular, we investigate the ex post
individual rationality, which is the strongest among the three
notions of individual rationality \cite{MASCOLELL95,GARG06}. In
the following theorem, we obtain a necessary and sufficient
condition for the ex post individual rationality of the BIC-B
protocol. Let $\hat \theta_i$ be the announced cost of a node $i$
and $\theta_i$ be the actual cost of that node.
\vspace{0.2cm}\\
\noindent
\textbf{Theorem 2:} The {\em BIC-B} protocol is {\em ex post
individual rational\/} if and only if
\begin{center} $\hat
\theta_i \le (\frac{n}{n-1})E[\theta_i]$, $\quad \forall i \in R$
\end{center}
where the $\hat \theta_i$ is the announced cost of the node $i$.
\vspace{0.1cm}\\
{\bf Proof:} See Appendix.

\subsection{An Implementation of the BIC-B Protocol}
In the previous sections, we have designed the BIC-B mechanism for
providing incentives to the nodes to participate in the broadcast
task. We have also proved certain important properties of the mechanism.
Here we provide a protocol implementation for the
BIC-B mechanism.

The protocol implementation is motivated by the emerging
technology of  hybrid ad hoc wireless networks, where
there are base stations that provide fixed infrastructure
for many network related functions. We propose that
additional functionality (which we call mediation functionality)
as described below be incorporated into
each base station.
Let us call this part of the base station as the {\em mediator\/}.
The mediator could be a part of any other network
infrastructure also.

The mediator elicits  the types from all the nodes, computes
the allocation and payments of the nodes, and announces the outcome.
We assume that all the nodes can communicate with the mediator.

After receiving the messages from the mediator regarding the
payments, each node constructs an internal table as shown in Table
\ref{int-tab-paper}. Each row of the table corresponds to a node
in the network. Each row contains three fields of information: (a)
{\em Source ID}, which specifies the source node ID from which the
packet is originating, (b) {\em Node List}, which specifies the
set of nodes to which the packet needs to be forwarded, and (c)
{\em Payment}, which specifies the payment to be received or paid.
The following is the structure of internal table of a node:

\begin{table}[h]
\begin{center}
\begin{tabular}{ | c | c | c | }
        \hline
  {\bf Source ID} & {\bf Node List} & {\bf Payment}\\
\hline   &  &  \\
\hline   &  &  \\
\hline
\end{tabular}
\caption{Structure of Internal Table of a Node}
\label{int-tab-paper}
\end{center}
\end{table}

In view of the above, the BIC-B protocol can be implemented as follows.

{\em BIC-B Protocol:} If a node receives a broadcast packet, it
checks its internal table entry corresponding to the source ID of
the broadcast. Then it forwards the packet to the set of nodes
specified in the {\em Node List} field of the entry and receives
the payment as mentioned in the {\em Payment} field of the entry.
On the other hand, if the {\em Node List} field is empty, then the
node does not forward the packet to any node and makes a payment
as mentioned in the {\em Payment} field of the entry.

In the above protocol, all the payment information by the node is
communicated directly to the mediator which takes care of
all the book keeping.


\section{Performance of the BIC-B Protocol} \label{simulation}
In this section, we show the efficacy of the proposed {\em BIC-B}
protocol for the {\em ICB} problem. In our simulation experiments,
we compare the performance of the {\em BIC-B} protocol with that
of the {\em Dominant Strategy Incentive Compatible Broadcast
(DSIC-B)\/} protocol  \cite{SURI06a}, \cite{SURITECH07}.

\subsection{Simulation Model} The {\em DSIC-B\/} protocol is
based  on {\em dominant strategy equilibrium} of the underlying
game and the {\em BIC-B} protocol is based on the {\em Bayesian
Nash equilibrium} of the underlying game. Since every dominant
strategy equilibrium is also a Bayesian Nash equilibrium, but not
vice-versa, we first find a dominant strategy equilibrium of the
underlying game and compute the payments to the nodes using the
{\em DSIC-B\/} and the {\em BIC-B} protocols \cite{SURI06a}.

We work with a randomly generated graph of an ad hoc wireless
network with the number of nodes $n=5, \; 10, \; 15, \; 20, \; 25,
\; 30, \; 35, \; 40$. According to our network model presented in
Section II, the graph is node weighted where these weights are the
transit costs of the nodes chosen independently and uniformly from
a range $[1,50]$. After the nodes announce their types, we first
compute the least cost paths to all the nodes from the source node
and then construct an {\em SRBT}. Using the {\em SRBT}, we can
decide the set of routers. This fixes the allocation rule. Then we
compute payments to the nodes using the payment rule of the
appropriate broadcast protocol. In all our simulation experiments,
the results for the performance metrics are averages taken over
100 random instances.

\subsection{Simulation Results}
\begin{figure}[h]
\begin{center}
\rotatebox{270}{\scalebox{0.35}{\includegraphics{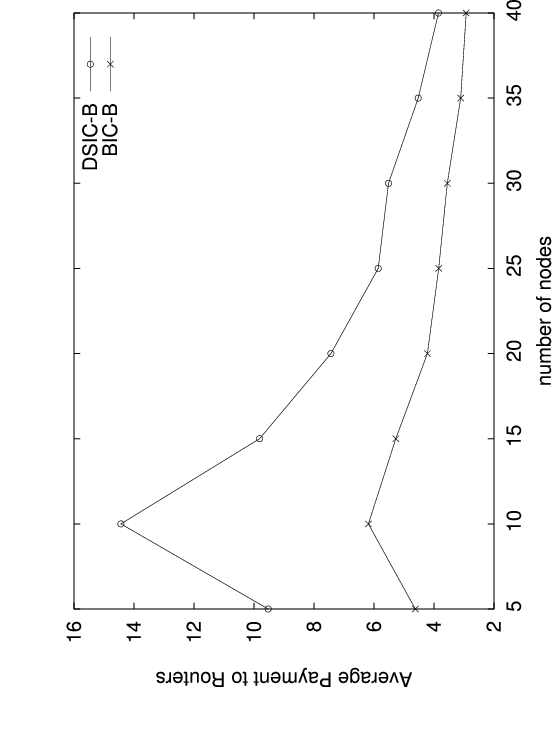}}}
\caption{\label{apr} Average payment to routers in the {\em
DSIC-B} and {\em BIC-B} protocols}
\end{center}
\end{figure}

We consider two performance metrics. The first metric is {\em
average payment to routers (APR)}. This specifies the payment on
an average to each router for forwarding the transit packets. The
graph in Figure \ref{apr} shows the comparison of the {\em BIC-B}
protocol and the {\em DSIC-B} protocol using APR. In the figure,
the lower curve corresponds to the {\em BIC-B} protocol. It is
clear from the figure that the {\em BIC-B} protocol performs
better than the {\em DSIC-B} protocol. This means that the system
wide payments made by the nodes to forward a broadcast packet is
less using the {\em BIC-B} protocol.


\input{epsf}

\begin{figure}[h]
\begin{center}
\rotatebox{270}{\scalebox{0.35}{\includegraphics{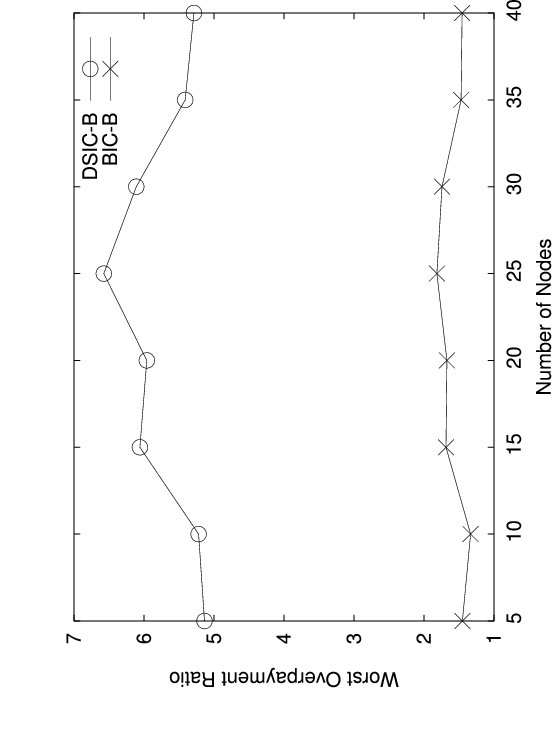}}}
\caption{\label{wor} Worst overpayment ratios for the {\em DSIC-B}
and {\em BIC-B} protocols}
\end{center}
\end{figure}

The second performance metric, we follow, is the {\em worst
overpayment ratio (WOR)}. We first define {\em overpayment ratio}
as the ratio of payment made by a node to its least cost path
value from the source node $s$. Now, we can define WOR as the
maximum over the overpayment ratios of all the nodes in the
network. That is,
\begin{center}
WOR = ${\rm max}_{i \in N} \frac{\rm payment \; made \; by \; node
\; i}{cost \; of \; path \; from \; source \; of \; broadcast \;
to \; node \; i}$
\end{center}
where {\em cost of path from source of broadcast to node $i$} is
the sum of forwarding costs of the nodes that lie on the path from
the source of broadcast node to the node $i$. Ideally we expect
this ratio to be $1$. We compare the WOR of {\em BIC-B} protocol
and the {\em DSIC-B} protocol in Figure \ref{wor}. We note that
the lower curve corresponds to the {\em BIC-B} protocol in the
figure. From the graph, it is easy to see that the worst
overpayment ratio in the network is higher using the {\em DSIC-B}
protocol than {\em BIC-B} protocol. WOR conveys the following
significant information. When a node receives a packet from a
router, then clearly the payment made by the receiver node to the
router is higher than the value of its least cost path, since it
has to give incentives to the router to make it reveal the true
incurred cost. If we take a ratio of the payment to the value of
least cost path, from the Figure \ref{wor}, this ratio is less
than 2 times over all the nodes using the {\em BIC-B} protocol and
it is higher than 5 times over all the nodes using the {\em
DSIC-B} protocol. This says that nodes end up with very high
payments, using {\em DSIC-B} protocol, than actually what their
value of the corresponding least cost path.


\input{epsf}

\section{Conclusions and Future Work}

We considered the incentive compatible broadcast (ICB) problem in
ad hoc wireless networks with selfish nodes. We proposed an
incentive based broadcast protocol that satisfies Bayesian
incentive compatibility and minimizes the incentive budgets
required by the individual nodes. The proposed protocol, {\em
BIC-B\/}, also satisfies budget balance. We also derived a
necessary and sufficient condition for the ex-post individual
rationality of the BIC-B protocol. We showed that the {\em
BIC-B\/} protocol exhibits superior performance when compared to a
dominant strategy incentive compatible solution to the problem.
Thus in this paper, we have addressed the ICB problem by proposing
an incentive mechanism and design of the BIC-B protocol.

While designing
incentive based protocols, we note that the complete solution
includes (a) design of the incentive mechanism (b) design of a
protocol which implements the incentive mechanism and (c)
addressing any problems that may arise, such as the cheating
problem. Our main contribution in this paper is in designing an
incentive mechanism. We have also briefly addressed the design of
a protocol in the current version of the paper. To address the
cheating, we need to invoke cryptographic techniques. There is
some literature available on the use of cryptographic techniques
to implement protocols, for example \cite{LILI06}.

In terms of mechanism design, it would be interesting to explore
optimal broadcast mechanisms in the {\em Myerson\/} sense
\cite{MASCOLELL95,DGARG08}. These are mechanisms that minimize the
incentive budgets subject to Bayesian incentive compatibility and
individual rationality.

An important problem that could be explored is design of
Bayesian incentive compatible protocols for unicast and multicast
problems. Existing game theoretic approaches to unicast and multicast
are all based on VCG mechanisms and have the usual limitations
associated with the use of VCG mechanisms.

Also, it is important to address certain practical
issues that arise in the implementation of these mechanisms as
part of standard protocols.
For example, the payment computation is performed in a centralized way in the
BIC-B protocol. It would be interesting to design a distributed
algorithm for this problem that could help deploy the BIC-B protocol
in the real world.
\section{Acknowledgment}
The research in this paper is partially supported by a research on
Algorithmic Mechanism Design for Complex Game Theoretic Problems
funded by the Office of Naval Research (Grant No. N0014-06-1-0994), Arlington, VA,
USA. We wish to thank Dr Shantanu Das, Program Manager,
Communications and Networking, for the encouragement and support.



\bibliographystyle{IEEEbib}
\bibliography{icb-jsac}

\begin{IEEEbiography}[{\includegraphics[width=1in,height=1.25in,clip,keepaspectratio]{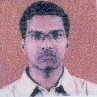}}]{N. Rama Suri}
received M.Sc.(Engg.) degree in computer science, in 2006, from
the Indian Institute of Science, Bangalore, India. He is currently
a Ph.D. scholar in Indian Institute of Science, Bangalore, India.
His current research interests include game theory, mechanism
design, ad hoc wireless networks, social networks, electronic
commerce, WWW. He is recipient of Microsoft Research India Ph.D.
fellowship for the duration 2007-2011.
\end{IEEEbiography}

\begin{IEEEbiography}[{\includegraphics[width=1in,height=1.25in,clip,keepaspectratio]{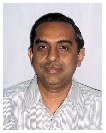}}]{Dr. Y. Narahari}
is currently a Professor at the Department of Computer Science and
Automation, Indian Institute of Science, Bangalore. He completed
his Ph.D. at the same Department in 1988, with his Doctoral
Dissertation on Petri Nets winning the Best Thesis Award for
Electrical Sciences at the Indian Institute of Science. His
current research focuses on the use of game theory and mechanism
design in network economics problems. He is currently completing a
research monograph entitled Emerging Game Theoretic Problems in
Network Economics and Mechanism Design Solutions, to be published
by Springer, London. He  has spent sabbaticals at the
Massachusetts Institute of Technology, Cambridge, Mass, USA, in
1992 and at the National Institute of Standards and Technology,
Gaithersburg, MD, USA, in 1997. He is currently on the editorial
board of IEEE Transactions on Systems, Man \& Cybernetics (Part A)
and IEEE Transactions on Automation Science \& Engineering (where
he is a Senior Editor). Dr. Narahari is a Fellow of the IEEE, a
fellow of the Indian National Academy of Engineering, a fellow of
the Indian National  Academy of Sciences, and a Homi Bhabha
Research Fellow. He has been involved in several high impact
collaborative research projects with General Motors R \& D, Intel,
and Infosys Technologies.
\end{IEEEbiography}

{\small
\begin{appendix}

\subsection{Proof of Lemma 1}
Note that the types of the
nodes are statistically independent according to the current
network model. For this reason, it does not matter even though we
take the expectation, in the above expression, with respect to
$\theta_i$ or $\theta_j$, where $i \in R$ and $j \notin R$. Now it
is easy to prove the lemma. Let the nodes in the set $R$ be
indexed by the set $\{1, 2, ..., r\}$, where $r = |R|$. Now for
any $j \notin R$ and $i \in R$,

\begin{tabular}{l l}
& $ E_{{{\theta}}_{-j}}\left[\sum_{l \in R, \: l \not =
i}\theta_{l}\right]$ \\
= & $\int...\int \left[\sum_{l \in R, \: l \not =
i}\theta_{l}\right] \; q(x_1)...q(x_{j-1})q(x_{j+1})...q(x_{n}) $\\
  & $\quad \quad \quad d(x_1)...d(x_{j-1})d(x_{j+1})...d(x_{n})$ \\
= & $\int...\int \left[\sum_{l \in R, \: l \not =
i}\theta_{l}\right] \; q(x_1)...q(x_{r}) \quad d(x_1)...d(x_{r})$\\
& (since types are statistically independent) \\
= & $\int...\int \left[\sum_{l \in R, \: l \not =
i}\theta_{l}\right] \; q(x_1)...q(x_{i-1})q(x_{i+1})...q(x_{r})$\\
& $\quad \quad \quad d(x_1)...d(x_{i-1})d(x_{i+1})...d(x_{r})$ \\
& (since $\left[\sum_{l \in R, \: l \not = i}\theta_{l}\right]$ does not include $\theta_i$) \\
= & $\int...\int \left[\sum_{l \in R, \: l \not =
i}\theta_{l}\right] \; q(x_1)...q(x_{i-1})q(x_{i+1})...q(x_{n})$\\
& $\quad \quad \quad d(x_1)...d(x_{i-1})d(x_{i+1})...d(x_{n})$ \\
& (since types are statistically independent) \\
= & $ E_{{{\theta}}_{-i}}\left[\sum_{l \in R, \: l \not =
i}\theta_{l}\right]$
\end{tabular}
\\\mbox{ }\hfill{\em Q.E.D.}

\subsection{Proof of Theorem 2} We show that the utility of each
router is non-negative after participating in the mechanism if and
only if the specified condition holds. This is nothing but proving
the ex post individual rationality of the nodes.

Let $f$ be the mechanism for the ICB problem. Now for ex post
individual rationality to hold for the router nodes,
\begin{center}
$u_i(f(\hat \theta),\hat \theta_i) \ge 0$, $\forall \hat \theta
\in \Theta, \quad \forall i \in R$.
\end{center}
where $\hat \theta$ is the vector of announcements of the costs
(or types) of the nodes. We also call $\hat \theta$ as the
announced cost (or type) profile of the nodes.

Now we characterize the utility of each node $i \in R$. All nodes
are the intended recipients of the packet(s) in a broadcast in the
network. We know, from Observation 1, that the payment made by a
non-router node is equivalent to $t_m(\hat \theta)$ for any $m
\notin R$. Since the routers are also intended recipients, they
also need to pay this amount. But, actually the routers do not pay
this amount and hence it is credited to their utility. Now we
have, $\forall \hat \theta \in \Theta$,

\begin{equation}
u_i(f(\hat \theta),\hat \theta_i)  =  v_i(k(\hat \theta_i,\hat
\theta_{-i})) - t_m(\hat \theta) + t_i(\hat \theta)
\label{utility_function}
\end{equation}
We have $t_m(\hat \theta) < 0$ for any $m \notin R$, from Lemma 2.
Hence this term appears with negative sign in the expression
(\ref{utility_function}). Now substituting the expression
(\ref{payment_rule}) in (\ref{utility_function}) and rearranging
the terms, we get $\forall \hat \theta \in \Theta$,

\begin{center}
\begin{tabular}{l c l}
$u_i(f(\hat \theta),\hat \theta_i)$ & = & $- \hat \theta_i $ +
$E_{{{\theta}}_{-m}}\left[\sum_{l \in R}\theta_{l}\right]$ $- E_{{{\theta}}_{-i}}\left[\sum_{l \in R, \: l \neq i}\theta_{l}\right] $\\
 &  & $+ \left(\frac{1}{n-1}\right) \sum_{j \neq i} E_{{{\theta}}_{-j}}\left[\sum_{l \in R, \: l \not = j} \theta_{l}\right]
 $\\
        &   & $- \left(\frac{1}{n-1}\right) \sum_{j \neq m} E_{{{\theta}}_{-j}}\left[\sum_{l \in R, \: l \not = j} \theta_{l}\right] $
\end{tabular}
\end{center}
By canceling out appropriate the terms, we get $\forall \hat
\theta \in \Theta$,
\begin{center}
\begin{tabular}{l c l}
$u_i(f(\hat \theta),\hat \theta_i)$ & = & $- \hat \theta_i $ +
$E_{{{\theta}}_{-m}}\left[\sum_{l \in R}\theta_{l}\right] $ $ -
E_{{{\theta}}_{-i}}\left[\sum_{l \in R, \: l \neq
i}\theta_{l}\right] $\\
&  & $+ \left(\frac{1}{n-1}\right)
E_{{{\theta}}_{-m}}\left[\sum_{l \in R} \theta_{l}\right]$ \\
              &   & $-  \left(\frac{1}{n-1}\right) E_{{{\theta}}_{-i}}\left[\sum_{l \in R, \: l \not = i} \theta_{l}\right] $ \\

              & = & $- \hat \theta_i $ +  $E_{{{\theta}}_{-m}}\left[\sum_{l \in R}\theta_{l}\right]$ $ - E_{{{\theta}}_{-m}}\left[\sum_{l \in R, \: l \neq
              i}\theta_{l}\right]$\\
              &   & $+ \left(\frac{1}{n-1}\right) E_{{{\theta}}_{-m}}\left[\sum_{l \in R}
              \theta_{l}\right]$\\
              &   & $- \left(\frac{1}{n-1}\right) E_{{{\theta}}_{-m}}\left[\sum_{l \in R, \: l \not = i} \theta_{l}\right] $ \\
              &   & (consequence of Lemma 1) \\

              & = & $- \hat \theta_i $ +  $E_{{{\theta}}_{-m}}\left[\sum_{l \in R}\theta_{l}\right] $ $- E_{{{\theta}}_{-m}}\left[\sum_{l \in R}\theta_{l}
              \right]$ \\
              &   & $+ E_{{{\theta}}_{-m}}\left[ \theta_i \right] + \left(\frac{1}{n-1}\right) E_{{{\theta}}_{-m}}\left[\sum_{l \in R}
              \theta_{l}\right]$\\
              &   & $  -  \left(\frac{1}{n-1}\right) E_{{{\theta}}_{-m}}\left[\sum_{l \in R} \theta_{l}\right]$
              $ + \left(\frac{1}{n-1}\right) E_{{{\theta}}_{-m}}\left[ \theta_i  \right] $\\
              &   & (by expanding the Expectation terms) \\

              & = & $- \hat \theta_i + E[\theta_i]+ \left(\frac{1}{n-1}\right) E[\theta_i] $\\

              & = & $\frac{nE[\theta_i] -(n-1)\hat \theta_i}{n-1}$
\end{tabular}
\end{center}
For the {\em BIC-B} protocol to be ex post individually rational,
\begin{center}
$u_i(f(\hat \theta),\hat \theta_i) \ge 0$, \quad $\forall \hat
\theta \in \Theta, \quad \forall i \in R$
\end{center}
From the above characterization of utility function, we get
\begin{center}
$\frac{nE[\theta_i] -(n-1)\hat \theta_i}{n-1} \ge 0, \quad \forall i \in R$ \\
\end{center}
This implies,
\begin{center}
$\hat \theta_i \le (\frac{n}{n-1})E[\theta_i], \quad \forall i \in
R$.
\end{center}
Furthermore, it is easy to see that each of above arguments are
reversible and hence the specified condition is  necessary and
sufficient for \emph{BIC-B} protocol to satisfy ex post individual
rationality.
\\\mbox{ }\hfill{\em Q.E.D.}
\end{appendix}}

\end{document}